%% file: main_7_arXiv.tex
\documentclass[10pt,
			   aps,
			   prb,
			   twocolumn, 
			   nofootinbib, 
			   showkeys,           
			   superscriptaddress  
			   ]{revtex4-1}

\usepackage{amssymb,amsmath}
\usepackage{cmap}
\usepackage{graphicx}
\usepackage{color}
\usepackage{hyperref}
\usepackage{verbatim}
\usepackage{blindtext}
\usepackage{listings}
\usepackage{booktabs}
\usepackage{multirow}
\hypersetup{
    bookmarks=true,          
    unicode=false,           
    pdftoolbar=true,         
    pdfmenubar=true,         
    pdffitwindow=false,      
    pdfstartview={FitH},     
    pdfauthor={Ivan Toftul}, 
    colorlinks=true,         
    linkcolor=blue,          
    citecolor=red,           
}
\usepackage{dsfont}

\usepackage{tabularx}
\AtBeginDocument{
	\heavyrulewidth=.08em
	\lightrulewidth=.05em
	\cmidrulewidth=.03em
	\belowrulesep=.65ex
	\belowbottomsep=0pt
	\aboverulesep=.4ex
	\abovetopsep=0pt
	\cmidrulesep=\doublerulesep
	\cmidrulekern=.5em
	\defaultaddspace=.5em
}

\graphicspath{{fig/}}

\renewcommand{\vec}[1]{\boldsymbol{\mathbf{#1}}}
\newcommand{\ve}[1]{\boldsymbol{\mathbf{#1}}}
\renewcommand{\t}[1]{\text{#1}}
\renewcommand{\Re}{\operatorname{Re}}
\renewcommand{\Im}{\operatorname{Im}}
\renewcommand{\d}{\partial}

\newcommand{\ten}[1]{\hat{\vec{#1}}}
\newcommand{\G}{\ten{G}}

\newcommand{\veps}{\varepsilon}

\newcommand{\f}{\t{f}}
\newcommand{\p}{\t{p}}
\newcommand{\m}{\t{m}}
\newcommand{\rf}{R_{\f}}
\newcommand{\rp}{R_{\p}}
\newcommand{\half}{\frac{1}{2}}

\begin{document}

\title{Self-trapped nanoparticle binding via waveguide mode 
}

\newcommand{\affilITMO}{ITMO University, Birzhevaya liniya 14, 199034 St.-Petersburg, Russia}

\author{I.D. Toftul}
\affiliation{\affilITMO}
\author{D.F. Kornovan}
\affiliation{\affilITMO}
\author{M.I. Petrov}
\email{m.petrov@metalab.ifmo.ru}
\affiliation{\affilITMO}

\date{\today}

\begin{abstract}
In this paper, we study a stable optomechanical system based on a nanoparticle chain coupled to a waveguide mode. Under the plane wave excitation the nanoparticles form a stable self-organized periodic chain array along the direction of the waveguide  through the transverse binding effect. We show that owing to the long-range interaction between the nanoparticles the trapping potential for each nanoparticle in the chain increases linearly with the system size making the formation of long chain more favourable. We propose that this effect can be observed with an optical nanofiber which is a versatile platform for achieveing optical binding of atoms and nanoparticles. Our calculations show that  binding energy for two nanoparticles is in the range of $9\div13~kT$  reaching the value of  100 $kT$ when the size of the chain is increased to 20 nanoparticles that makes potential experimental observation of  the effect possible. We also suggest the geometry of the two counter propagating plane waves excitation, which will allow trapping the nanoparticles close to the optical nanofiber providing efficient interaction between the nanoparticles and the nanofiber.  
\end{abstract}


\maketitle

\textit{Introduction.}
The optical manipulation provides unique opportunities for  controlling  micro and nanoobjects at the remarkable level of precision, which has find applications in nanophysics \cite{GrierDavidG2003}, biochemistry \cite{Poeggel2015}  and biology \cite{WyattShieldsIv2015},  and  allows testing the fundamental limits of quantum physics \cite{Frimmer2016}. One of the main applications of optomechanics is related to manipulation of large atomic and particles ensembles   providing a reliable platform for studying many-body physics. By modulating the spatial distribution of the  optical fields, one can form two-dimensional\cite{Bloch2005,Gross2017} and three-dimensional\cite{GrierDavidG2003,Kumar2018,Barredo2018} trapping potential for ultracold atoms and nanoparticles in the free-space. An alternative method of structuring the large ensembles bases on the self-assembly approach \cite{Cizmr2010}. The rescattering of the optical fields by the trapped objects results in effective dipole-dipole interaction which can lead to structurizing of the  large ensembles. The weak optical interactions can be enhanced and modified with auxiliary photonic structures such as metamaterials\cite{Bogdanov2015a} and metausurfaces\cite{Ivinskaya2018}, plasmonic structures\cite{Ivinskaya2017, Marago2013}, as well as dielectric nanofibers\cite{Frawley2014}. The latter one presents a versatile platform \cite{Daly2015} for studying light interaction with nanoparticles \cite{Maimaiti2016,Joos2018} and atoms \cite{Vetsch2010c, Corzo2019} placed close to its surface. Utilization of a single mode  long-range dipole-dipole interactions provided by waveguiding systems  has already been suggested for self-organization of atoms and nanoparticles in waveguiding systems \cite{Chang2013,Holzmann2014,Bykov2018}.  In this paper, we predict a novel feature of a waveguide driven self-assembled  nanoparticle systems, which manifests itself in the deepening of the trapping potential for each nanoparticle with the increase of the number of particles trapped in the system. Thus, increasing the system's size results in its higher optomechanical stability, ensuring that the self-assembled system will be as large as possible. 
 
\begin{figure}
	\includegraphics[width=1\linewidth]{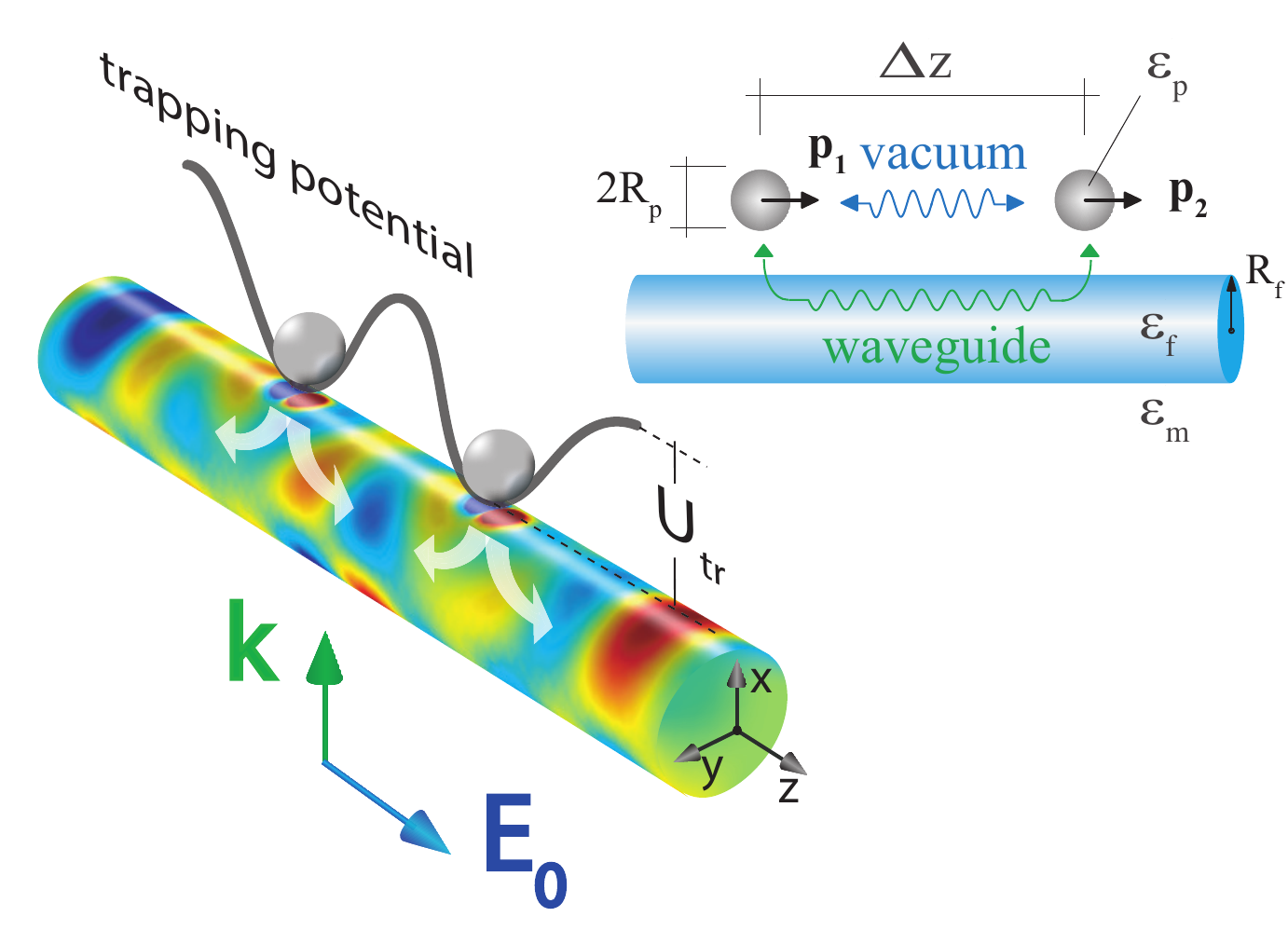}
	\caption{The proposed system configuration which allows  achieving  nanoparticles binding close the nanofiber under transversal  plane illumination.}
	\label{fig:mainfig}
\end{figure}
 We propose a particular geometry of nanoparticle chain placed close to the nanofiber and illuminated by a plane wave propagating in the free space perpendicularly to the fiber axis as it is shown in Fig.\ref{fig:mainfig}. Such geometry allows to take the advantage of the transverse optical binding effect \cite{Burns1989, Dholakia2010}.  The binding occurs due to the interference of the fields scattered by the nanoparticles, and it has been applied for self-organization of nanoparticle ensembles under the external laser illumination \cite{Chaumet2001, Chvatal2015,Demergis2012}, including the interference of surface plasmon polariton modes \cite{Kostina2019}. In our work, the nanofiber modes allow for accumulation of long-range interactions between distant nanoparticles, which results in the increasing  particles stiffness with the growth of  the nanoparticle chain length. Moreover, in the particular geometry of nanofiber binding, we also suggest a method for trapping the nanoparticles in the radial direction close to the fiber surface by using two counter propagating plane waves and taking the advantage of nanofiber focusing effect \cite{LeKien2009}. Thus, we suggest the geometry of the system that allows for immediate testing of the reported effect within the particular experimental setup with use of an optical nanofiber.

\begin{figure}[ht]
	\includegraphics[width=0.9\linewidth]{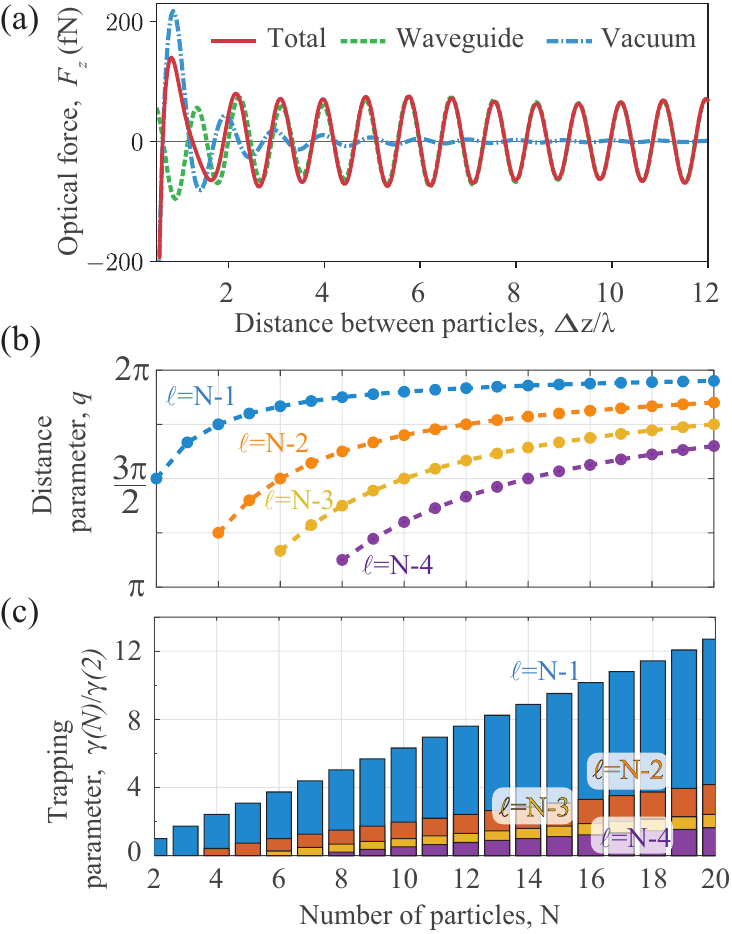} 
	\caption{{(a)} Longitudinal optical force $F_z$ acting on one of  two particles as a function of distance along the fiber axis $\Delta z$ for a SM regime. Red solid line shows optical force considering interaction both through fiber and  free space ($\ten{G}_s + \ten{G}_0$), green dashed line shows only interaction through the fiber ($\ten{G}_s$) and blue dash-and-dot line shows only free space interaction ($\ten{G}_0$). {(b)} First three branches of the solution for the average distance between the two closest particles with  respect to the number of particles in a chain. (c) Trapping parameter with respect to the number of of particles in a chain. Value is normalized by the trapping parameter for the two particles $\gamma(2)$.}
	\label{fig:SM}
\end{figure}
{\it Single-mode nanoparticle binding.} The force acting on a single dipole particle and averaged over the period of the incident wave  is given by \cite{Chaumet2000} $	\vec F  = \frac{1}{2} \sum_i\Re p^*_i \nabla E_{\t{loc,}i}$, where the sum is taken over the Cartesian components of the  dipole moment ${\bf p}$ and local field ${\bf E}_{loc}$. The latter one contains the  incident plane wave field $\vec{E}_{\t{inc}}$, the field  scattered  by the nanofiber $\vec{E}_{\text{sc}}$, and the field scattered by other nanoparticles $\vec{E}_{p}$. The dipole moment of $n$-th particle then will be defined through the local field strength $\ve p_i=\alpha_0 \ve E_{\t{loc}} (\ve r_i)=\alpha_0  (\ve E_{0}(\ve r_i) + \ve E_p(\ve r_i))$, where we defined $\ve E_0=\ve E_{\t{inc}}+\ve E_{\t{sc}}$ as the external field, and $\alpha_0$ is the exact dipole polarizability given by the Mie theory \cite{doyle1989optical}. The dipole field $\ve E_p$ is the field generated by other nanoparticles and can be expressed via Green's function formalism. For instance, the field generated by the $i$-th particle at the positions of $j$-th particle has the form $\ve E_{p,ij}=k_0^2/\veps_0\G(\ve r_i, \ve r_j) \ve p_j$, where $\ten{G} = \ten{G}_0 + \ten{G}_\t{s}$ is the total Green's tensor which consists of two parts: free-space $\ten{G}_0$ and scattered $\ten{G}_\t{s}$, which appears du to the presence of the nanofiber   (see Appendix \ref{app:greenTensor}). Here $k_0$ is the vacuum wavenumber and $\veps_0$ is the vacuum permittivity.

\textit{Binding of nanoparticles with a single-mode nanofiber.} The size of nanofiber can be chosen in such that it supports only a single guiding HE$_{11}$ mode \cite{marcuse1972light} (SM-single mode regime). In this case one can expect almost periodicity in the interaction strength between the nanoparticles with the interparticle distance $\Delta z$. Indeed, in Fig.~\ref{fig:SM}~a) the optical force between two nanoparticles positioned close to the nanofiber is shown as a function of the distance between the particles. The contributions of vacuum and nanofiber  interaction channels  are extracted by proper choosing the Green's function. One can see  that the force has a well-pronounced periodic character, which allows  forming of a stable configuration of nanoparticle chain consisting of arbitrary number of particles \cite{Holzmann2014, Holzmann2016}. 
 
In a single mode regime the Green's  function \eqref{eq:Gs11}  of the waveguide can be reduced to $\ten{G}_{s}^{\t{wg}} (\vec{r}; \vec{r}') = \ten{\mathcal{G}}_{s}^{\t{wg}} (\rho, \varphi; \rho', \varphi') e^{i \beta |\Delta z|}$, and $\ten{G}_{s}^{\t{wg}} (\vec{r}; \vec{r})$ is purely imaginary at the origin  for any waveguiding mode \cite{Yao2010}. Here, we  neglect the contribution of leaky and evanescent modes \cite{marcuse1972light} as they decay significantly at long distances.    In the field of  a plane wave incident normally to the nanofiber and polarized along the $z$-axis (TM polarized) as shown in Fig.~\ref{fig:mainfig} the dipole moments will be aligned preferably along  the nanofiber axis, thus, having dominant $z$-component of the dipole moment $\vec{p}_i \approx \vec{n}_{z} p_i = \vec{n}_z \alpha_{\t{eff},z z} E_{0,z}$, (see Appendix \ref{app:alphaGeneral} for the details). TM excitation allows to suppress the vacuum interaction channel as the dipole emission along the nanofiber axis is weak. The force acting on a particle with number $n$ can be estimated as:  
\begin{align}
	F^{\t{SM}}_n =  \frac{|p|^2 k_0^3 \beta}{2\varepsilon_0} &\Im(g_{zz}) \times \nonumber \\
	&  {\sum_{j \neq n}^{N} \cos (\beta |z_n - z_j|)\operatorname{sign}(j - n)}.
	\label{eq:force_n}
\end{align}
Here $N$ is the total number of particles in the chain, and we introduced the coupling constant $g_{zz}(\rho)=\ten{\mathcal{G}}_{\t{s}, z z}^{\t{wg}} (\rho, \varphi; \rho, \varphi)/k_0$, which depends only on the radial distance to the  nanoparticle center in the geometry shown in Fig.~\ref{fig:mainfig}.
The system within the considered approximations has a stable equidistant configuration where separation between the neighboring nanoparticles is constant \cite{Holzmann2016}. In order to find it, one needs also to estimate the stiffness parameter $\kappa$, which determines the strength of the restoring force $F_{z}=-\kappa (z-z_{0})$ acting on a single particle close to the equilibrium position $z_0$. This approach is valid as non-conservative part of the binding optical force  is negligible. The stable configuration of nanoparticles is observed if the separation distance between the neighbouring nanoparticles satisfies two conditions: (i) $F_{n,z} \propto \sum_{j \neq n}^{N} \cos (q |n - j|) \operatorname{sign}(j - n) = 0$  and (ii) $\kappa_n \propto - \sum_{j \neq n}^{N} \sin (q |n - j|) >0$ for all particles. Here $q=\beta \Delta z$ is the distance parameter and $\Delta z$ is the distance between the neighbouring nanoparticles.  After taking the sum in  Eq.~\eqref{eq:force_n}, the  first condition provides us with the expression for the equidistant solution $Nq/2=\pi/2+\pi \ell$, where $\ell$ is an integer. 

The stiffness of the $n$-th trap $\kappa_n = - \d_{z_n} F_z$ in the chain of $N$ particles can be estimated as follows:

\begin{equation}
	\kappa^{\t{SM}}_n = - \frac{|p|^2 k_0^3 \beta^2}{2\varepsilon_0} \Im (g_{zz}) {\sum_{j \neq n}^{N} \sin (q |n - j|)},
	\label{eq:kappa_n}
\end{equation}
and the summation is taken in order to account for the interaction with  all nanoparticles in the chain. The stability condition requires that $\kappa_n$ should be positive for any particle in the chain. The analytical solution of the algebraic system shows that there exists a set of    stable configurations. The fundamental solution with the smalles value of the interparticle distance $q$  corresponds to $\ell=N-1$ and has the distance parameter \cite{Holzmann2014} $q_1=2\pi-\pi/N$ (see blue line in Fig.~\ref{fig:SM} b)). Moreover, the stiffness parameter $\kappa_n=\kappa(N)$ is the same for any particle in the chain and increases with the growth of total number of particles in the chain as $\kappa(N)\sim \cot(\pi/2N)$, which for  $N\gg1$ provides the linear increase of the stiffness $\kappa(N)\sim N$ as shown in Fig.~\ref{fig:SM} c). 
Other stable equidistant configurations correspond to other values of $\ell$ and have larger distance parameter $q=2\pi+(\ell-N)\pi/N$, $\ell=N-1, N-2 \dots $ and $l\geq N/2$, as shown in Fig.~\ref{fig:SM} b) for $\ell=N-2, N-3, N-4$. The $\kappa_n$ values for these solutions also demonstrate the linear growth with $N$, however, with a smaller slope than for $\ell=N-1$ (see Fig.~\ref{fig:SM} c)). 

\begin{table}[h]
	\centering
	\caption{Proposed parameters of the system.}
	\label{tab:prm}
	\resizebox{\linewidth}{!}{%
		\begin{tabular}{@{}cccccc@{}}
			\toprule
			Regime & \begin{tabular}[c]{@{}c@{}}Fiber radius\\ $\rf$\end{tabular} & \begin{tabular}[c]{@{}c@{}}Wavelength\\ $\lambda_0 = 2\pi c/ \omega$\end{tabular} & \begin{tabular}[c]{@{}c@{}}Particle radius\\ $\rp$\end{tabular} & $\rf/\lambda_0$      & $V$-number  \\ \midrule
			SM     & 300 nm                                                       & \multirow{2}{*}{1064 nm}                                        & \multirow{2}{*}{150 nm}                                         & 0.28               & 1.860             \\  
			MM     & 1000 nm                                                       &                                                                &                                                                 & 0.94                & 6.201             \\ 
			\midrule
			\multicolumn{2}{c}{Media permittivity, $\varepsilon_{\text{m}}$}                                   & \multicolumn{2}{c}{Particle permittivity, $\varepsilon_{\text{p}}$}                                                                                              & \multicolumn{2}{c}{Fiber permittivity, $\varepsilon_{\text{f}}$} \\ 
			\multicolumn{2}{c}{1.0}                                              & \multicolumn{2}{c}{2.5}                                                                                                          & \multicolumn{2}{c}{2.1025}             \\ 
			\midrule
			\multicolumn{2}{c}{Distance to the fiber,  $d$}                                   & \multicolumn{2}{c}{Pump power, $P$}                                                                                              & \multicolumn{2}{c}{Pump field magnitude, $E_0$} \\ 
			\multicolumn{2}{c}{SM: $ 45 \text{ nm}$ \quad MM: $ 50 \text{ nm}$}                                              & \multicolumn{2}{c}{$200\t{ mW}$}                                                                                                          & \multicolumn{2}{c}{$2.45\cdot10^6\ \t{V}/\t{m}$}             \\ \bottomrule
		\end{tabular}%
	}
\end{table} 

\begin{figure}[htbp]
	\centering
	\includegraphics[width=\linewidth]{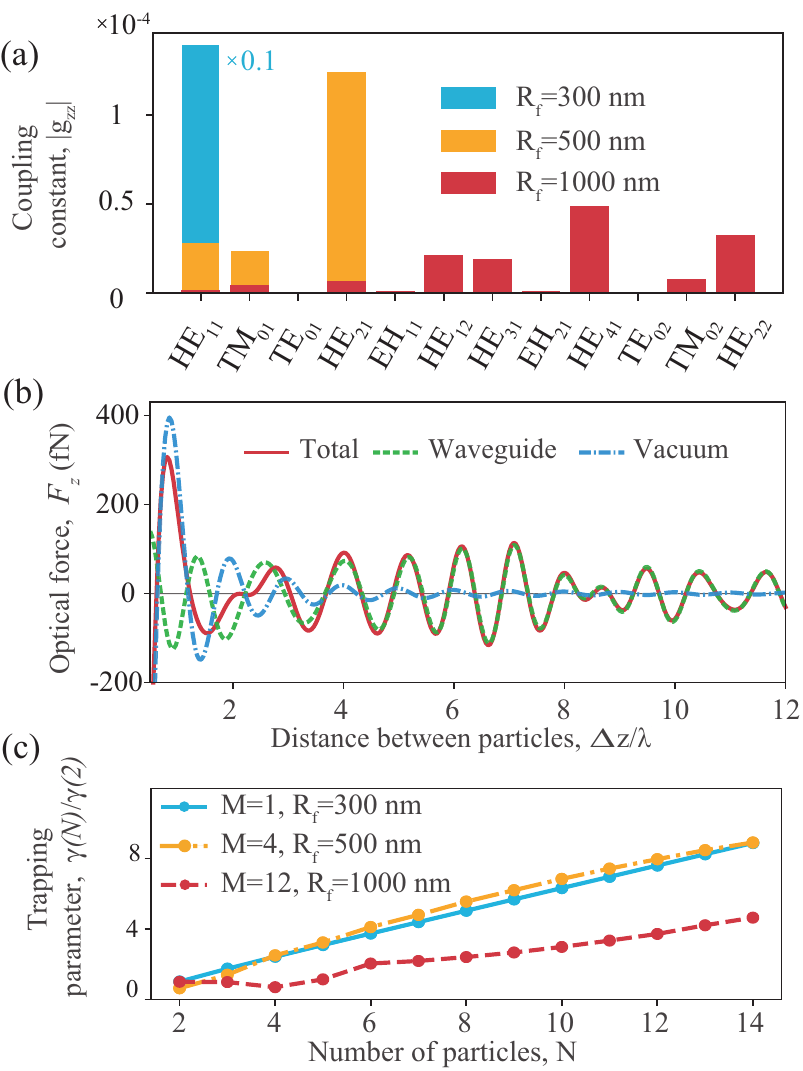}
	\caption{ (a) The amplitude of the coupling constant $g_{zz}$ for different modes in SM and MM regimes for three different radiuses $R_f=300, 500, 1000$ nm. In the MM regime   the biggest contribution give HE$_{21}$ and $HE_{41}$ modes for 500 nm and 1000 nm radiuses correspondingly. One can note that for given excitation geometry TE modes are not excited in the nanofiber. (b) Longitudinal optical force $F_z$ acting on one of the particles as a function of distance along fiber axis $\Delta z$ for MM regime $R_f=1000$ nm. Red solid line shows optical force considering interaction both through fiber and through free space, green dashed line shows  interaction through the fiber only, and blue dash-and-dot line shows only free space interaction.(c) The  trapping potential of nanoparticles in a equidistant chain states as a function of number of particles in M and MM regimes. The results are obtained with the optimization procedure. }
	\label{fig:MM}
\end{figure}

{\it Stability of trapping.} Increase of the stiffness of each nanoparticle's trap basically leads to increased stability of the chain, which can be interpreted in terms of  the trapping parameter $\gamma_{\t{tr}} = U_{\t{tr}} / kT$, where $U_{\t{tr}}$ is the trapping potential separating the stable and unstable positions of each  particle in the chain. It can be expressed as $U_{\t{tr}}(N) = \kappa(N) \Delta z^2 /2$, where $\Delta z\approx \pi/\beta$. The trapping potential for the fundamental configuration for $\ell=N-1$ then can be estimated in the single mode approximation as follows:

\begin{gather}
	\gamma_{\t{tr}}^{\t{SM}} (N) = \frac{\pi^2 k_0^3 |p|^2}{4 kT \veps_0} \Im(g_{zz}^{wg})\cot\left(\dfrac{\pi}{2N}\right)\propto N\ \text{for}\ N\gg 1.
\label{eq:trapp_N}
\end{gather}
  This expression is one of the main results of the paper, showing that the stability of the considered system increases linearly  with the growth of nanoparticle number in the chain. This basically means that the self-ordering of nanoparticles in the longer chain will be more preferential, and in fact is only limited by the width of the exciting laser beam and light intensity as $\gamma_{\t{tr}}\sim|p|^2\sim |E_0|^2$.

In order to support our analytical results and estimate the achievable values of trapping potential, we  used the full  model, describing interacting dielectric nanoparticles placed close to a nanofiber. We took into  account the plane wave rescattering on the nanofiber, the nanoparticles self-polarization effect due to the nanofiber presence, as well a nanoparticle cross polarization effects. For the set of parameters close to the experimental ones\cite{nieddu2016} and summarized in Table 1, the calculations give us the  estimation of the binding parameter for two nanoparticles $\gamma(2)\approx 9$ at room temperature, which  is a promising value for the potential experimental applications. Moreover, according to Fig.~\ref{fig:SM} in the  chain consisting of $N=20$ nanoparticles in the fundamental configuration one can expect $\gamma(20)\approx 110$, i.e. the trapping potential can be two-orders of magnitude higher than $kT$.

{\it Nanoparticle binding  in multi-mode regime.} With the increase of nanofiber radius the number of the waveguide modes starts to rapidly increase which significantly changes the picture of nanoparticles interaction. The coupling constants of  each mode are depicted in Fig.~\ref{fig:MM} (a). One can see that the higher modes give the bigger contribution to the coupling constant as their field penetration outside the waveguide is stronger. The simultaneous excitation of different modes provides aperiodic interaction potential between two particles. Our computational model allows for a full modelling of multi-mode (MM) interaction between the nanoparticles, and the computed optical binding force is shown in Fig.~\ref{fig:MM}~(b) for the parameters specified in Table~\ref{tab:prm}. Our estimations of the the trapping parameter for  MM regime  give the value of $\gamma^{\t{MM}}_{\t{tr}}(2) \approx 13$ for the room temperature, which is higher than in a single mode regime due to larger number of modes and their stronger field penetration outside the waveguide~\cite{kumar2015interaction}. 

Despite the aperiodic interaction, one still can expect the effect of self-induced organization of nanoparticles via transverse binding.   In the MM regime Eq.~\eqref{eq:kappa_n} will gain another sum over many interaction channels corresponding to different  waveguide modes:
\begin{eqnarray}
\label{eq:kappa_n_MM}
	\kappa^{\t{MM}}_n &=& - \half |p|^2 \frac{k_0^3}{\varepsilon_0}  \times
	\\ && \nonumber
	\sum_{\mu = 1}^{M}\beta_\mu^2 \Im \left\{{g}_{ z z}^{\beta_\mu} \right\} \Im \sum_{j = 1}^{N} e^{i \beta_\mu |z_n - z_j|},
\end{eqnarray}
where $\beta_\mu = \beta_{\t{HE}_{11}}, \beta_{\t{TM}_{01}}, ...$ are the propagation constants of the allowed modes (see the dispersion curve in Fig.~\ref{fig:forBinding_dispersion} in Supplementary materials), and $M$ defines the number of the allowed waveguide modes. The stable configuration of the nanoparticle chain can be found through the maximization of \eqref{eq:kappa_n_MM}. We applied a numerical optimization algorithm with proper constrains ($\kappa_n>0$, $F_{n,z} = 0$ for any $n$) to identify the nanoparticles configuration and the stiffness of the trap. The optimization procedure started by a configuration of ordered chain separated with the distance  $\Delta z \beta_{\t{max}} = 4\pi-\pi/N$, where $\beta_{\t{max}}$ is the propagation constant corresponding to the dominant mode among  all the excited ones (HE$_{21}$ and HE$_{41}$ for $R_f=500$ nm  and $R_f=1000$ nm respectively). 
The final result after optimization procedure is the average stiffness $\left\langle \kappa \right\rangle$ and corresponding noramlized trapping parameter  presented in Fig.~\ref{fig:MM} c). One can see that the system demonstrates the stable configuration, which averaged trapping parameter increases linearly with the size $N$ similar to the SM case.


\begin{figure}
	\centering
	\includegraphics[width=\linewidth]{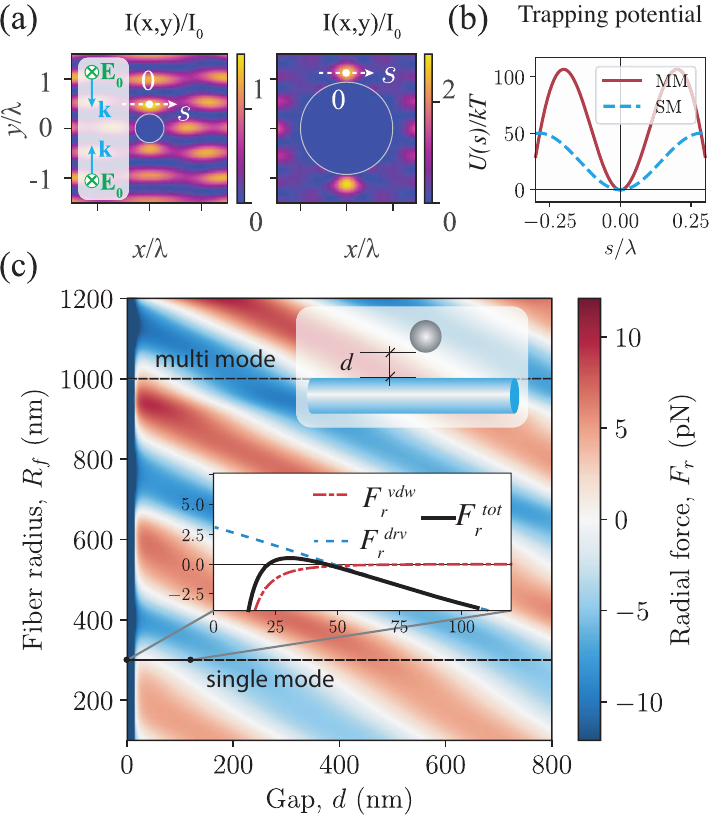}
	\caption{
		{(a)} Normalized intensity of the total electric field for the two geometry. 
		{(b)} Potential energy of the transverse trap along $x$ axis normalized by $kT$.
		{(c)} Total radial force as a function of two parameters: the fiber radius and the gap between fiber surface and particle's surface. Two horizontal white lines represent fiber radius choices: $\rf^{\t{SM}} = 300$~nm and $\rf^{\t{MM}} = 1000$~nm. On the inset plot there is the force decomposition to optical force and van-der-Vaals force. For the two selected radii we have $d^{\t{SM}} = 45$~nm and $d^{\t{MM}} = 50$~nm equilibrium distances. Other parameters are specified in Table~\ref{tab:prm}.}
	\label{fig:2beams}
\end{figure}
{\it Radial and azimuthal binding of nanoparticles.} Finally, it is worth speculating about the potential mechanisms of trapping of the nanoparticles close to the nanofiber surface.
Illumination of the nanofiber by a plane wave forms an interference pattern in the vicinity of the nanofiber \cite{LeKien2009}, which can act as a trapping potential for nanoparticles. However, a single beam illumination  also provides a strong optical pressure force acting on nanoparticles, which prevents effective trapping in the radial direction. We suggest a geometry with   two counter propagating interfering beams fully compensating the optical pressure force and enabling strong binding of nanoparticles close to the nanofiber surface. The formed potential trap provides both radial and azimuthal stability of the nanoparticles. In   Fig~\ref{fig:2beams}~a) the field intensity distribution normalized by the intensity of the plane wave is shown around the SM and MM nanofibers providing stability of nanoparticles in the trap in the transverse direction along the $s$-axis. The trapping parameter cross section is depicted Fig.~\ref{fig:2beams} (b) demonstrating  the values of 50 and 100 for SM and MM nanofiber, respectively.

The radial stability of the particles is studied  in Fig.~\ref{fig:2beams} (c), where the radial force acting on the particle  is shown as a function of the gap between fiber and particle surface $d$ and fiber radius $\rf$. Total radial force also includes the contribution from the van-der-Waals attractive force~\cite{gu1999van} $\vec{F}^{\t{vdW}}$ , along with the electromagnetic force $\vec{F}^{\t{em}}$. The white regions in the  2D  map correspond to the regions of zero optical force and, thus, the regions of radially stable configurations where the force changes sign from positive to negative with the increase of the gap. The two dashed lines denote the SM and MM nanofiber radiuses. In Fig.~\ref{fig:2beams} (c) inset the cross section of the total radial force is shown for SM regime, showing that a stable point at the gap distance of 45 nm can be achieved. Finally, one should note that by adding a phase difference between the up- and down-propagating  interfering beams one can  gradually modify the radial trapping potential and finely tune the position of the radially stable points (see Appendix~\ref{app:beams}).        



\textit{Conclusions.} In this work, we proposed that an optomechanical  systems coupled through a waveguide mode can demonstrate stable configurations, which stability will be increased with growth of the  particles number in the system. This counterintuitive result is provided by a one-dimensional character of the interaction meaning that  under stochastic self-assembly process formation of larger systems will be preferable to the appearance of smaller ones. We theoretically predict that  a  nanoparticle chain located close to an optical nanofiber can demonstrate the transverse optical binding effect with the trapping potential for two dipole nanoparticles being  one order of magnitude higher than a thermal energy at ambient conditions. Moreover, in the case of longer chains the trapping potential can be  increases with the number of particle, reaching a value of 110 $kT$  for nanoparticles chain consisting of 20 particles. We also propose an excitation geometry based on two counter-propagating beams, in which one can achieve stable radial and azimuthal trapping, which will form a two-dimensional potential locating the nanoparticles chain close to nanofiber, and making the proposed effect potentially observable in the experiment.  Finally, the result of our paper can be extended to other self-organized one- and two-dimensional systems, where the stability of the systems will grow with its size. 
  

 

---------------------------------------------------------------------

\section*{Acknowledgments}

The authors are grateful for the discussions to Prof. Sile Nic Chormaic and Dr. Viet Giang Truong. M.P. and D.K. have been supported by the Foundation for the Advancement of Theoretical Physics and Mathematics "Basis."

\bibliography{FiberBinding_bib_2}

\appendix 
\input{FiberBinding_appendix}

\end{document}

%% file: FiberBinding_appendix.tex
\section{\label{app:greenTensor}Fiber Green's tensor}

The classical electromagnetic Green's tensor of our system can be found from the vector Helmholtz equation:  

\begin{eqnarray}
\left[- \frac{\omega^2}{c^2}\veps(\mathbf{r},\omega) + \mathbf{\nabla} \times \mathbf{\nabla} \times  \right] \mathbf{G}(\mathbf{r},\mathbf{r}^\prime,\omega)=\mathbf{I}\delta(\mathbf{r}-\mathbf{r}^\prime),
\label{A1}
\end{eqnarray}
where $\veps(\mathbf{r},\omega)$ is the complex dielectric function and $\mathbf{I}$ is the unit dyad. In our case, we consider a dielectic cylindrical waveguide of radius $\rho_c$ and dielectric permittivity $\veps$ being constant inside the cylinder. To find the solution we apply the scattering superposition method \cite{Chew1999, Tai1994}, which allows to expand the Green's tensor into the homogeneous and inhomogeneous terms:
\begin{equation}
\mathbf{G}(\mathbf{r},\mathbf{r}^\prime,\omega) = \mathbf{G}_0(\mathbf{r},\mathbf{r}^\prime,\omega) + \mathbf{G}_s(\mathbf{r},\mathbf{r}^\prime,\omega).
\end{equation}

As soon as we consider dielectric particles in the vicinity of the waveguide, so that $\mathbf{r}, \mathbf{r}^\prime$ are outside the cylinder, the homogeneous term is always present and describes the field directly generated at the field point $\mathbf{r}$ by the source placed at the point $\mathbf{r}^\prime$. This term can be obtained analytically from the Green tensor written in cartesian coordinates using the transformation from cartesian to cylindrical coordinates $\mathbf{S}(\varphi)\mathbf{G}_{0}^{Cart}(\mathbf{r},\mathbf{r}^\prime,\omega)\mathbf{S}^T(\varphi')$, where $\mathbf{G}_{0}^{Cart}$ has an analytic expression \cite{novotny2012principles} and is given by 
\begin{eqnarray}
\mathbf{G}_{0}^{Cart}(\mathbf{r}, \mathbf{r^\prime}, \omega) = \left( \mathbf{I} + \frac{1}{k^2}\mathbf{\nabla} \otimes \mathbf{\nabla}\right)G_0(\mathbf{r},\mathbf{r}^\prime,\omega),
\end{eqnarray}
here $G_0(\mathbf{r},\mathbf{r}^\prime,\omega) = e^{ik|\vec{r} - \vec{r}'|} / 4 \pi |\vec{r} - \vec{r}'|$ is the  Green's function of the scalar Helmholtz equation.

The scattering term can be calculated via the integral representation of the homogeneous part. To obtain this representation we apply the method of vectorial wave function (VWF) explained in details in Ref.~\cite{Chew1999, Tai1994}, here we cover only the basic ideas and provide the final expressions.
To find the solution of the vector Helmholtz equation \eqref{A1} we introduce the scalar Helmholts equation and  the solution of this equation in the cylindrical coordinates:
\begin{eqnarray}
&& \nabla^2\phi(\mathbf{k},\mathbf{r}) + k^2\phi(\mathbf{k},\mathbf{r}) =0, 
\nonumber\\
&& \phi_n(k_{z},\mathbf{r})=J_n(k_\rho \rho)e^{in\varphi+ik_zz},
\end{eqnarray}
here $J_n(x)$ is the Bessel function of the first kind, $\mathbf{r} = (\rho,\varphi,z)$ are the cylindrical coordinates and $k_\rho$, $k_z$ are the projections of the wavevector $\mathbf{k}$.
The solution of the vector Helmholtz equation may be written in terms of the following vector wavefunctions:
\begin{eqnarray}
\mathbf{M}_n(k_z,\mathbf{r}) &=& \mathbf{\nabla} \times [\phi_n(k_z,\mathbf{r})\mathbf{e_z}] 
\nonumber\\
\mathbf{N}_n(k_z,\mathbf{r}) &=& 
\frac{1}{k}\mathbf{\nabla } \times \mathbf{M}_n(k_z,\mathbf{r}) 
\end{eqnarray}
where $\mathbf{e_z}$ is the so-called pilot vector, the unit vector pointing in the $z$ direction. These VWFs $\mathbf{M_n}(k_z, \mathbf{r})$, $\mathbf{N_n}(k_z, \mathbf{r})$ correspond to $TE/TM$ modes of the field.

One can show  \cite{Chew1999} that the homogeneous part of the Green's function can be expanded in terms of these vector wavefunction in the following way:
\begin{eqnarray}
\mathbf{G}_{0}(\mathbf{r},\mathbf{r^\prime}, \omega)=-\dfrac{\mathbf{e_{\rho}e_{\rho}}}{k_0^2} \delta (\mathbf{r}-\mathbf{r^\prime}) +\nonumber\\
+\dfrac{i}{8\pi}\sum_{n= - \infty}^{\infty} \int\limits_{-\infty}^{\infty} \frac{dk_z}{k_{0\rho}^2} \mathbf{F}_n (k_z, \mathbf{r}, \mathbf{r^\prime})
\end{eqnarray}
and the $\mathbf{F}_n (k_z, \mathbf{r}, \mathbf{s})$ function is given by
\begin{equation}
\begin{cases}
\mathbf{M}_n^{(1)} (k_z, \mathbf{r})\overline{\mathbf{M}}_n (k_z, \mathbf{r^\prime}) + \mathbf{N}_n^{(1)} (k_z, \mathbf{r})\overline{\mathbf{N}}_n (k_z, \mathbf{r^\prime})&  \\
\mathbf{M}_n (k_z, \mathbf{r})\overline{\mathbf{M}}_n^{(1)} (k_z, \mathbf{r^\prime}) + \mathbf{N}_n (k_z, \mathbf{r})\overline{\mathbf{N}}_n^{(1)} (k_z, \mathbf{r^\prime})& 
\end{cases}
\end{equation}
here the first line holds for $\rho_r>\rho_{r^\prime}$ while the second one for $\rho_r<\rho_{r^\prime}$, and $k_0={\omega}/{c}$, $k_{0\rho }= \sqrt{k_0^2-k_z^2}$ and the superscript $(1)$ in vector wave functions denotes that the Bessel function of the first kind $J_n(k_\rho \rho)$ should be replaced with the Hankel function of the first kind $H^{(1)}_n(k_\rho \rho)$. Here we provide the explicit form of VWF:

\begin{eqnarray}
\mathbf{M}_n(k_z,\mathbf{r}) &=& 
\begin{pmatrix}
\frac{in}{\rho} J_n(k_{0\rho } \rho)\\
- k_{0\rho } (J_n(k_{0\rho } \rho))'\\
0
\end{pmatrix} e^{i n \varphi + i k_z z},
\nonumber\\ 
\mathbf{N}_n(k_z,\mathbf{r}) &=& 
\begin{pmatrix}
\frac{ik_z k_{0\rho}}{k} (J_n(k_{0\rho} \rho))'\\
-\frac{n k_z}{\rho k} J_n (k_{0\rho } \rho)\\
\frac {k_{0\rho}^2}{k} J_n (k_{0\rho} \rho)
\end{pmatrix} e^{i n \varphi + i k_z z} 
\nonumber\\
\overline{\mathbf{M}}_n(k_z,\mathbf{r^\prime}) &=& 
\begin{pmatrix}
-\frac{in}{\rho^\prime} J_n(k_{0\rho } \rho^\prime)\\
- k_{0\rho } (J_n(k_{0\rho } \rho^\prime))'\\
0
\end{pmatrix}^T e^{- i n \varphi^\prime - i k_z z^\prime},
\nonumber\\ 
\overline{\mathbf{N}}_n(k_z,\mathbf{r^\prime}) &=& 
\begin{pmatrix}
-\frac{ik_z k_{0\rho }}{k} (J_n(k_{0\rho } \rho^\prime))'\\
-\frac{n k_z}{\rho^\prime k} J_n (k_{0\rho } \rho^\prime)\\
\frac {k_{0\rho }^2}{k} J_n (k_{0\rho } \rho^\prime)
\end{pmatrix}^T e^{- i n \varphi^\prime - i k_z z^\prime}\raisetag{10\baselineskip} \nonumber\\
{} 
\end{eqnarray}
where $J_n(k_\rho \rho)'$ means derivative with respect to the dimensionless argument.

Now having the integral representation of the homogeneous term of the Green's function, we can construct the scattering term in a similar fashion. Let us denote the medium outside the dielectric cylinder as $1$ and the medium inside as $2$. The particular form of the Green's tensor depends on the position of a source point $\mathbf{r^\prime }$: whether it is inside or outside the cylinder. As soon as we are interested in a situation, when both source and receiver are outside the cylinder and in the latter we consider only the second case. Thus, the total Green's tensor can written as:
\begin{eqnarray}
\begin{cases}
\mathbf{G}^{11}(\mathbf{r},\mathbf{r^\prime},\omega) = \mathbf{G}^{11}_0(\mathbf{r},\mathbf{r^\prime},\omega) + \mathbf{G}^{11}_s(\mathbf{r},\mathbf{r^\prime},\omega), \\
\mathbf{G}^{21}(\mathbf{r},\mathbf{r^\prime},\omega) = \mathbf{G}^{21}_s(\mathbf{r},\mathbf{r^\prime},\omega), 
\end{cases}
\end{eqnarray}
here two superscripts denote position of the receiver the source point respectively and the two scattering parts of the Green's tensor has the following form:
\begin{eqnarray}
\mathbf{G}_{s}^{11}(\mathbf{r,r^\prime,\omega}) &=&
\frac{i}{8 \pi} \sum_{n= - \infty}^{\infty} \int\limits_{-\infty}^{\infty} \frac{dk_z}{k_{\rho1}^2} \mathbf{F}^{11 (1)}_{\mathbf{M};n, 1}(k_z,\mathbf{r})\overline{\mathbf{M}}_{n,1}^{(1)}(k_z,\mathbf{r^\prime}) 
\nonumber \\  
&+&\mathbf{F}^{11 (1)}_{\mathbf{N};n,1}(k_z,\mathbf{r})\overline{\mathbf{N}}_{n,1}^{(1)}(k_z,\mathbf{r^\prime}) ,
\nonumber\\
\mathbf{F}^{11 (1)}_{\mathbf{M};n,1}(k_z,\mathbf{r}) &=& R^{11}_{MM}\mathbf{M} ^{(1)}_{n,1}( k_z, \mathbf{r})+R^{11}_{NM} \mathbf{N}_{n,1}^{(1)} (k_z,\mathbf{r}) ,
\nonumber\\
\mathbf{F}^{11 (1)}_{\mathbf{N};n,1}(k_z,\mathbf{r}) &=& R^{11}_{MN} \mathbf{M} ^{(1)}_{n,1}(k_z, \mathbf{r})+R^{11}_{NN} \mathbf{N}_{n,1}^{(1)} (k_z,\mathbf{r}).
\label{eq:Gs11}\\
\mathbf{G}_{s}^{21}(\mathbf{r,r^\prime,\omega}) &=&
\frac{i}{8 \pi} \sum_{n= - \infty}^{\infty} \int\limits_{-\infty}^{\infty}  \frac{dk_z}{k_{\rho1}^2} \mathbf{F}^{21}_{\mathbf{M};n,2}(k_z,\mathbf{r})\overline{\mathbf{M}}_{n,1}^{(1)}(k_z,\mathbf{r^\prime}) 
\nonumber \\ 
&+& \mathbf{F}^{21}_{\mathbf{N};n,1}(k_z,\mathbf{r})\overline{\mathbf{N}}_{n,1}^{(1)}(k_z,\mathbf{r^\prime}) ,
\nonumber\\
\mathbf{F}^{21}_{\mathbf{M};n,2}(k_z,\mathbf{r}) &=& R^{21}_{MM} \mathbf{M}_{n,2}(k_z, \mathbf{r})+R^{21}_{NM} \mathbf{N}_{n,2} (k_z,\mathbf{r}) ,
\nonumber\\
\mathbf{F}^{21}_{\mathbf{N};n,2}(k_z,\mathbf{r}) &=& R^{21}_{MN} \mathbf{M}_{n,2}(k_z, \mathbf{r})+R^{21}_{NN} \mathbf{N}_{n,2} (k_z,\mathbf{r}), \label{eq:Gs21}
\end{eqnarray}
here the scattering Fresnel coefficients $R_{AB}^{ij}$ are introduced and the second subscript in the VWFs denotes that $k$ and $k_\rho$ should be replaced with their values inside the corresponding media $k_i=\sqrt{\veps_i(\mathbf{r},\omega)} k_0$, $k_{\rho i}=\sqrt[]{k_i^2 - k_z^2}$ and also $k_{0\rho}$ becomes $k_{\rho i}$. We should notice that unlike the case of the homogeneous term, here we have products of $\mathbf{M}$ and $\mathbf{N}$, which is due to the fact that the normal modes in our case have hybrid nature.

The form of the Fresnel coefficients mentioned above can be found by imposing the boundary conditions on the Green's tensor at the surface of the cylinder 
\begin{eqnarray}
{\begin{cases}
	\mathbf{e}_{\rho} \times [\mathbf{G}^{11}(\mathbf{r},\mathbf{r^\prime},\omega) - \mathbf{G}^{21}(\mathbf{r},\mathbf{r^\prime},\omega) ]|_{\rho_r = \rho_c} = 0, 
	\\
	\mathbf{e}_{\rho} \times \mathbf{ \nabla_r } \times [\mathbf{G}^{11}(\mathbf{r},\mathbf{r^\prime},\omega) - \mathbf{G}^{21}(\mathbf{r},\mathbf{r^\prime},\omega)]|_{\rho_r = \rho_c} = 0.
	\end{cases}}\nonumber\\
{}
\end{eqnarray}
Solving for this, we can find the Fresnel coefficients $R_{AB}^{ij}$ and, finally, construct the scattering part of the Green's tensor $\mathbf{G}_s(\mathbf{r},\mathbf{r}^\prime,\omega)$. We provide the explicit expressions for the Fresnel coefficients below:

\begin{widetext} 
	\begin{eqnarray}
	& DT(k_z) = -\left( \dfrac{1}{k_{\rho2}^2} - \dfrac{1}{k_{\rho1}^2} \right)^2 k_z^2 n^2 + \left( \dfrac{(J_n(k_{\rho2}\rho_c))^\prime}{k_{\rho2}J_n(k_{\rho2}\rho_c)} - \dfrac{(H^{(1)}_n(k_{\rho1}\rho_c))^\prime}{k_{\rho1}H^{(1)}_n(k_{\rho1}\rho_c)}\right) \times \nonumber\\
	& \left( \dfrac{(J_n(k_{\rho2}\rho_c))^\prime k_2^2}{k_{\rho 2}J_n(k_{\rho2}\rho_c)} - \dfrac{(H^{(1)}_n(k_{\rho1}\rho_c))^\prime k_1^2}{k_{\rho1}H_n^{(1)}(k_{\rho1}\rho_c)}\right) \rho_c^2 \nonumber\\
	& R_{MM}^{11}(k_z) = \dfrac{J_n(k_{\rho1}\rho_c)}{H^{(1)}_n(k_{\rho1}\rho_c)} \Bigg[ \left( \dfrac{1}{k_{\rho2}^2} - \dfrac{1}{k_{\rho1}^2} \right)^2 k_z^2 n^2 - \left( \dfrac{(J_n(k_{\rho2}\rho_c))^\prime}{k_{\rho2}J_n(k_{\rho2}\rho_c)} - \dfrac{(J_n(k_{\rho1}\rho_c))^\prime}{k_{\rho1}J_n(k_{\rho1}\rho_c)} \right) \times \nonumber\\
	& \left( \dfrac{(J_n(k_{\rho2}\rho_c))^\prime k_2^2}{k_{\rho2}J_n(k_{\rho2}\rho_c)} - \dfrac{(H^{(1)}_n(k_{\rho1}\rho_c))^\prime k_1^2}{k_{\rho1}H^{(1)}_n(k_{\rho1}\rho_c)} \right) \rho_c^2\Bigg] \dfrac{1}{DT(k_z)} \nonumber\\
	& R_{NM}^{11}(k_z) = \dfrac{J_n(k_{\rho1}\rho_c)}{H_n^{(1)}(k_{\rho1}\rho_c)}\dfrac{1}{k_{\rho1}} \left( \dfrac{1}{k_{\rho1}^2} - \dfrac{1}{k_{\rho2}^2}\right) \left( \dfrac{(J_n(k_{\rho1}\rho_c))^\prime}{J_n(k_{\rho1}\rho_c)} - \dfrac{(H^{(1)}_n(k_{\rho1}\rho_c))^\prime}{H_n^{(1)}(k_{\rho1}\rho_c)}\right) \dfrac{k_1 k_z n \rho_c}{DT(k_z)} \nonumber\\
	& R_{MN}^{11}(k_z) = R_{NM}^{11} \nonumber\\
	&R_{NN}^{11}(k_z) = \dfrac{J_n(k_{\rho1}\rho_c)}{H^{(1)}_n(k_{\rho1}\rho_c)} \Bigg[ \left( \dfrac{1}{k_{\rho2}^2} - \dfrac{1}{k_{\rho1}^2} \right)^2 k_z^2 n^2 - \left( \dfrac{(J_n(k_{\rho2}\rho_c))^\prime}{k_{\rho2}J_n(k_{\rho2}\rho_c)} - \dfrac{(H^{(1)}_n(k_{\rho1}\rho_c))^\prime}{k_{\rho1}H^{(1)}_n(k_{\rho1}\rho_c)} \right) \times \nonumber\\
	& \left( \dfrac{(J_n(k_{\rho_2}\rho_c))^\prime k_2^2}{k_{\rho2}J_n(k_{\rho2}\rho_c)} - \dfrac{(J_n(k_{\rho1}\rho_c))^\prime k_1^2}{k_{\rho1}J_n(k_{\rho1}\rho_c)}\right)\rho_c^2\Bigg]\dfrac{1}{DT(k_z)}
	\end{eqnarray}
\end{widetext}

\section{Effective polarizability. General case}
\label{app:alphaGeneral}

The effective polarizability of a nanoparticle placed in the vicinity of a nanofiber waveguide and accounted for the interaction with another particle can expressed as follows:
\begin{multline}
\ten{\alpha}_{\text{eff}}^{(i)} = \left[ \ten{I} - \left(\frac{k^2}{\varepsilon_0}\right)^2 \ten{\alpha}_s^{(i)} \G^{ij} \ten{\alpha}_s^{(j)} \G^{ji}\right]^{-1} \cdot \\
\cdot \ten{\alpha}_s^{(i)} \left( \ten{I} + \frac{k^2}{\varepsilon_0} \G^{ij} \ten{\alpha}_s^{(j)} \right),
\label{eq:alpha_eff}
\end{multline}
where $i = 1, 2$ and $j = 1 \cdot \delta_{i2} + 2 \cdot \delta_{i1}$, $\G = \G_0 + \G_s$ and
\begin{equation}
\ten{\alpha}_s^{(j)} = \alpha_0 \left( \ten{I} - \alpha_0 \frac{k^2}{\varepsilon_0} \G_s^{jj} \right)^{-1}.
\end{equation}

Here we consider the polarizability for the case when $\vec{E}_0 (\vec{r}_1) \neq \vec{E}_0 (\vec{r}_2)$. Hence, it follows that it is impossible to factor out the external field $\vec{E}_0$ to obtain expression as $\vec{p} = \ten{\alpha}_{\t{eff}} \vec{E}_0$. Yet we can still introduce the effective polarizability tensor if we convert it into an operator by introducing  a \textit{shifting operator}.  The obtained  operator of the effective polarizability will have a form: 
\begin{multline}
	\ten{\alpha}_{\text{eff}}^{(i)} = \left[ \ten{I} - \left(\frac{k^2}{\varepsilon_0}\right)^2 \ten{\alpha}_s^{(i)} \G^{ij} \ten{\alpha}_s^{(j)} \G^{ji}\right]^{-1} \cdot \\
	\cdot \ten{\alpha}_s^{(i)} \left( \ten{I} + \frac{k^2}{\varepsilon_0} \G^{ij} \ten{\alpha}_s^{(j)} e^{(\vec{r}_i - \vec{r}_j) \cdot \nabla} \right),
	\label{eq:alpha_eff_general}
\end{multline}
where $i = 1, 2$ and $j = 1 \cdot \delta_{i2} + 2 \cdot \delta_{i1}$, $\G = \G_0 + \G_s$, $e^{(\vec{r}_2 - \vec{r}_1) \cdot \nabla}$ is the shifting operator and
\begin{equation}
	\ten{\alpha}_s^{(j)} = \alpha_0 \left( \ten{I} - \alpha_0 \frac{k^2}{\varepsilon_0} \G_s^{jj} \right)^{-1}.
\end{equation}
Using \eqref{eq:alpha_eff_general} we can calculate dipole moment in a straightforward way as 
\begin{equation}
	\vec{p}_i = \ten{\alpha}^{(i)}_{\text{eff}} \vec{E}(\vec{r}_i).
\end{equation} 

We also need to note, that this expression is a generalization of what was used in this paper, as we considered that all the particles are placed in the same external field $\vec{E}_0 (\vec{r}_1) =\vec{E}_0 (\vec{r}_2)$

\section{\label{app:alphaAndMie}Polarizablity and Mie theory}

\textit{Electrostatic approximation} gives the following expression for the polarizability \cite{novotny2012principles}:
\begin{equation}
	\t{(SI)} \qquad \qquad \alpha_{\t{es}} = 4 \pi \varepsilon_0 \varepsilon_\m \rp^3 \frac{\veps_{\p} - \veps_{\m}}{\veps_{\p} + 2\veps_{\m}},
	\label{eq:standard}
\end{equation}
which connects dipole moment and external field as $\vec{p} = \alpha_{\t{es}} \vec{E}$. This solution may be refined by the radiation corrections \cite{le2013radiative, de1998point}:
\begin{equation}
	\alpha_{\text{rad}} = \frac{\alpha_{\t{es}}}{1 - i \alpha_{\t{es}} \frac{k^2}{\veps_0} \Im \G_0 (\vec{r}_0, \vec{r}_0)},
	\label{eq:rad_corr}
\end{equation}
where $\Im \G_0 (\vec{r}_0, \vec{r}_0) = k \sqrt{\veps_{\m}}/6\pi$.

The \textit{exact} expression for the electric polarizability of the sphere can be found using the Mie scattering theory \cite{doyle1989optical}:
\begin{equation}
	\t{(SI)} \qquad \qquad  \alpha_0 = 4\pi \varepsilon_0 \veps_\m \cdot i \frac{3 \rp^3}{2x^3} a_1 (x, m),
	\label{eq:alpha_mie}
\end{equation}
where $x =\sqrt{\varepsilon_m} k \rp$, $m =\sqrt{\varepsilon_{\text{p}}}/\sqrt{\varepsilon_{\text{m}}}$ and 
\begin{equation}
a_1(x,m) = \frac{m\psi_1 (mx) \left(\psi_1(x)\right)' - \psi_1 (x) \left(\psi_1(mx)\right)'}{m \psi_1 (mx) \left(\xi_1(x)\right)' - \xi_1(x) \left(\psi_1(mx)\right)'},
\end{equation}
where 
\begin{equation}
\begin{matrix}
\psi_1(z) = z j_1(z), \qquad \xi(z) = z h_1^{(1)}(z), \\ \\
h^{(1)}_1(z) = j_1(z) + i y_1 (z),
\end{matrix}
\end{equation}
where $j_1$ and $y_1$ are the spherical Bessel functions of the first and second kind correspondingly.


\section{Two beam trapping by phase tuning}
\label{app:beams}

\begin{figure}
	\centering
	\includegraphics[width=\linewidth]{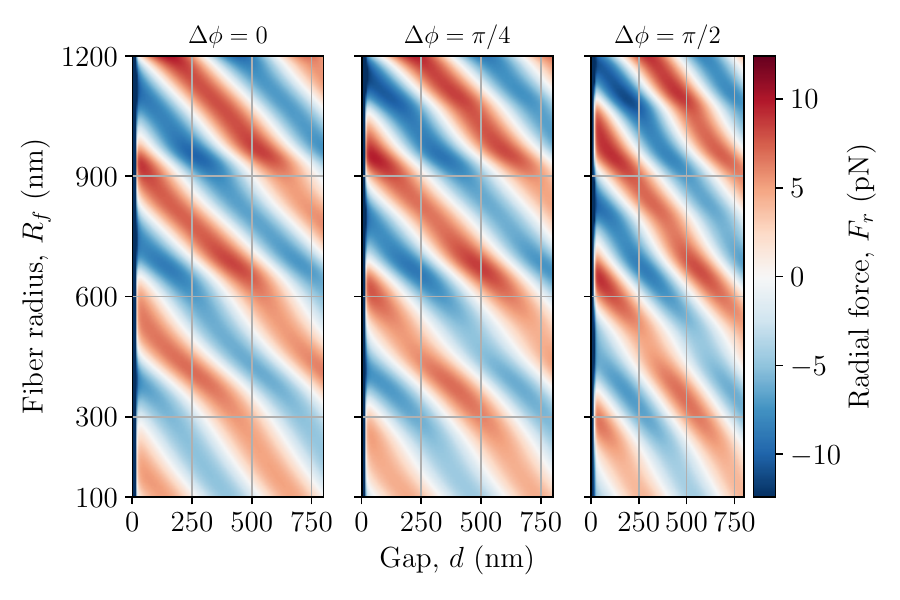}
	\caption{Total radial force on the particle for the two counterpropagating beams configuration as a function of two parameters: the fiber radius and the gap between fiber surface and particle's surface. Sequence of plots is shown for the various phase shift between two beams are plotted to show how equilibrium trapping distance can be tuned.}
	\label{fig:force2dmap_phase_shift}
\end{figure}

The usage of two beams instead of one give several advantages. 
At first, second beam supresses the scattering pressure which is capable to overcome gradient part of the force in the case of single beam transverse pump. Secondly, by the tuning relative phase shit between two counterpropagating perpendicular beams we can change position of nodes and antinodes and thus position of the radial trapping (see Fig.~\ref{fig:force2dmap_phase_shift}).

\section{Fiber dispersion}
\label{app:disp}

\begin{figure}
	\centering
	\includegraphics[width=\linewidth]{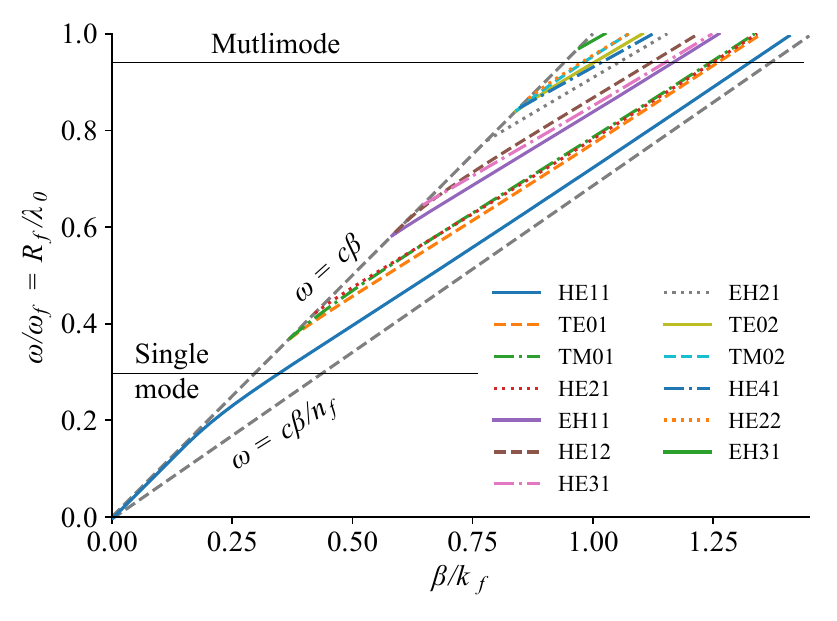}
	\caption{Fiber dispersion $\beta(\omega)$ for different guided modes. Here $k_{\f} = 2\pi/\rf$, $\omega_{\f} = 2\pi c/ \rf$, $n_{\f} = \sqrt{\veps_{\f}}$ and $c$ is the speed of light. All the parameters are taken according to Tab.~\ref{tab:prm}. For the MM regime there are 12 allowed modes.}
	\label{fig:forBinding_dispersion}
\end{figure}

Optical fiber represents itself as a cylindrical waveguide. Depending on its radius and relative permittivity, fiber can support different number of modes (branches in the dispersion equation $\beta = \beta(\omega)$~\cite{le2017higher, marcuse1972light}). Conventionally, there are 4 types of eigen modes: TE modes (transverse electric) with $E_z = 0$, TM modes (transverse magnetic) with $H_z=0$, and HE and EH modes (hybrid) with all 3 non-zero components of elecric and magnetic fields. The dispersion curve for the particular parameters of this paper is shown of Fig.~\ref{fig:forBinding_dispersion}.

